\DocumentMetadata{%
	lang=en-US,
	pdfstandard = A-3u,
	pdfversion = 1.7,
}

\documentclass[balance,nocopyright,upint,nolists]{asmejour}

\allowdisplaybreaks

\usepackage{graphicx}

\hypersetup{%
	pdfauthor={Jiachen Li, Edward Kim, Hanyu Zhu, Dongmei Chen, Wei Li},
	pdftitle={A Wearable ECG Device for Differentiating Hypertrophic Cardiomyopathy from Acquired Left Ventricular Hypertrophy},
	pdfkeywords={Hypertrophic cardiomyopathy, Left ventricular hypertrophy, Electrocardiogram, Wearable medical device, Cardiac screening},
	pdfsubject={Wearable ECG device for HCM diagnosis},
}

\JourName{Medical Devices}

\begin{document}

%%%%%%%%%%%%%%%%%%%%%%%%%%%%%%%%%%%%%%%%%%%%%%%%%%%%%%%%%%%%%%%%%%%%%%%%%%%%%%%%%
%%%%%%%%%%%%%%%%%%%%%%%%%%%%%%%%%%%%%%%%%%%%%%%%%%%%%%%%%%%%%%%%%%%%%%%%%%%%%%%%%
%%  AUTHORS
%%%%%%%%%%%%%%%%%%%%%%%%%%%%%%%%%%%%%%%%%%%%%%%%%%%%%%%%%%%%%%%%%%%%%%%%%%%%%%%%%
\SetAuthorBlock{Jiachen Li}{%
	Department of Mechanical Engineering,\\
	University of Texas at Austin,\\
	Austin, TX 78712 USA\\
	email: jiachenli@utexas.edu
}

\SetAuthorBlock{Hanyu Zhu}{%
	Department of Mechanical Engineering,\\
	University of Texas at Austin,\\
	Austin, TX 78712 USA\\
	email: zhuhanyu@utexas.edu
}

\SetAuthorBlock{Edward Kim}{%
	Department of Mechanical Engineering,\\
	University of Texas at Austin,\\
	Austin, TX 78712 USA\\
	email: edwardkim8775@utexas.edu
}

\SetAuthorBlock{Shihao Li}{%
	Department of Mechanical Engineering,\\
	University of Texas at Austin,\\
	Austin, TX 78712 USA\\
	email: shihaoli01301@utexas.edu
}

\SetAuthorBlock{Katherine Cavanaugh}{%
	Dell Medical School,\\
	University of Texas at Austin,\\
	Austin, TX 78712 USA\\
	email: katherine.cavanaugh@utexas.edu 
}

\SetAuthorBlock{Arpan Patel}{%
	Department of Cardiology,\\
	Division of Cardiac Electrophysiology,\\
	Austin, TX 78712 USA\\
	email: arpan.patel@ascension.org 
}

\SetAuthorBlock{Sovik De Sirkar}{%
	Department of Cardiology,\\
	Division of Cardiac Electrophysiology,\\
	Austin, TX 78712 USA\\
	email: sovik.de.sirkar@ascension.org 
}
\SetAuthorBlock{Mauricio Hong}{%
	Department of Cardiology,\\
	Division of Cardiac Electrophysiology,\\
	Austin, TX 78712 USA\\
	email: MHong@ascension.org 
}

\SetAuthorBlock{Wei Li\CorrespondingAuthor}{%
	Department of Mechanical Engineering,\\
	University of Texas at Austin,\\
	Austin, TX 78712 USA\\
	email: weiwli@austin.utexas.edu
}

\SetAuthorBlock{Dongmei Chen\CorrespondingAuthor}{%
	Department of Mechanical Engineering,\\
	University of Texas at Austin,\\
	Austin, TX 78712 USA\\
	email: dmchen@me.utexas.edu
}
%%%%%%%%%%%%%%%%%%%%%%%%%%%%%%%%%%%%%%%%%%%%%%%%%%%%%%%%%%%%%%%%%%%%%%%%%%%%%%%%%
%%  TITLE
%%%%%%%%%%%%%%%%%%%%%%%%%%%%%%%%%%%%%%%%%%%%%%%%%%%%%%%%%%%%%%%%%%%%%%%%%%%%%%%%%

\title{A Wearable ECG Device for Differentiating Hypertrophic Cardiomyopathy from Acquired Left Ventricular Hypertrophy}

%%%%%%%%%%%%%%%%%%%%%%%%%%%%%%%%%%%%%%%%%%%%%%%%%%%%%%%%%%%%%%%%%%%%%%%%%%%%%%%%%
%%  KEYWORDS AND ABSTRACT
%%%%%%%%%%%%%%%%%%%%%%%%%%%%%%%%%%%%%%%%%%%%%%%%%%%%%%%%%%%%%%%%%%%%%%%%%%%%%%%%%

\keywords{Hypertrophic cardiomyopathy, Left ventricular hypertrophy, Electrocardiogram, Wearable medical device, Cardiac screening}

\begin{abstract}
Hypertrophic Cardiomyopathy (HCM) is a genetic heart disease affecting approximately 1 in 500 people and is the leading cause of sudden cardiac death in young athletes. Current diagnostic methods---cardiovascular magnetic resonance (CMR), echocardiography, and genetic testing---are limited by high costs, operator dependency, or insufficient accuracy, while standard electrocardiogram (ECG) analysis cannot reliably distinguish HCM from acquired left ventricular hypertrophy (LVH). This paper presents a wearable ECG device paired with a classification algorithm that differentiates HCM from acquired LVH using ECG signals alone. The portable device integrates a 3-lead electrode system, an AD8232 signal conditioning module, an Arduino Nano 33 BLE microcontroller, and a lithium polymer battery. The algorithm extracts two quantitative indices---HCM Index~1 and HCM Index~2---from each heartbeat and classifies patients via dual statistical thresholds. Validation on 483 LVH patients (PhysioNet) and 29 HCM patients (digitized clinical records) yields 75.86\% sensitivity, 99.17\% specificity, and an F1-score of 80.00\%. Leave-one-out cross-validation confirms generalizability, with cross-validated sensitivity of 72.41\%, specificity of 98.96\%, and F1-score of 76.36\% (95\% confidence intervals reported). A digitization confound analysis demonstrates that the classification is driven by physiological cardiac features rather than data source artifacts. A simulated device acquisition chain analysis confirms that the wearable hardware's signal characteristics are compatible with the classification algorithm. The system offers a promising tool for affordable HCM screening in resource-limited settings.
\end{abstract}

\date{}

\maketitle

%%%%%%%%%%%%%%%%%%%%%%%%%%%%%%%%%%%%%%%%%%%%%%%%%%%%%%%%%%%%%%%%%%%%%%%%%%%%%%%%%
%%  INTRODUCTION
%%%%%%%%%%%%%%%%%%%%%%%%%%%%%%%%%%%%%%%%%%%%%%%%%%%%%%%%%%%%%%%%%%%%%%%%%%%%%%%%%

\section{Introduction}

Hypertrophic cardiomyopathy (HCM) is a genetic heart disease that causes abnormal thickening of the left ventricular muscle, affecting approximately 1 in 500 people worldwide~\cite{mckinney2024}. HCM is the leading cause of sudden cardiac death (SCD) in young athletes, accounting for nearly 40\% of SCD cases in this population~\cite{malhotra2017}. The disease is particularly dangerous because affected individuals are often asymptomatic until experiencing a life-threatening cardiac event, typically during or shortly after intense physical exertion. This makes early detection critical, especially for athletes participating in competitive sports where cardiovascular demands are high.

The disease is associated with mutations in genes encoding cardiac sarcomere proteins, including Myosin-binding protein C (MYBPC3), Beta-myosin heavy chain (MYH7), Troponin T (TNNT2), and Troponin I (TNNI3), with MYBPC3 being the most prevalent, occurring in 40--50\% of cases~\cite{lopes2024}. These genetic mutations cause disorganized arrangement of cardiac muscle cells (myocyte disarray) and abnormal collagen deposition, leading to the characteristic asymmetric thickening of the left ventricular wall. Unlike acquired left ventricular hypertrophy (LVH), which results from external factors such as chronic hypertension, aortic stenosis, or intensive athletic training and produces uniform, concentric thickening of the ventricular wall~\cite{shenasa2017}, HCM causes asymmetric hypertrophy that significantly increases the risk of life-threatening ventricular arrhythmias~\cite{hensley2015}. The distinction between HCM and acquired LVH is clinically crucial because the management strategies differ substantially: athletes with HCM are typically restricted from competitive sports, while those with physiological LVH from training (athlete's heart) can continue their activities safely.

Current diagnostic methods for HCM have significant limitations that prevent widespread screening. Cardiovascular magnetic resonance (CMR) imaging provides the most detailed cardiac visualization and is considered the gold standard for measuring wall thickness and detecting fibrosis, but the equipment costs up to one million dollars, making it inaccessible in many healthcare settings, particularly in rural areas and developing regions~\cite{pennell2004,murali2023}. Additionally, CMR requires specialized facilities and trained personnel to operate and interpret the results. Echocardiography is more affordable and widely available, but it is heavily operator-dependent, requiring specialized training that limits its reliability across different clinical settings~\cite{ryding2013,popescu2009}. Image quality can also be affected by patient body habitus, and the technique may miss subtle cases of HCM. Genetic testing offers high specificity when positive, but has variable diagnostic accuracy, ranging from 50--60\% for young patients with family history to only 30--40\% for patients without known family history~\cite{geneticalliance2009,bonaventura2021}. Furthermore, genetic testing is expensive, requires specialized laboratories, and results can take weeks to obtain. These limitations highlight the critical need for an accurate, accessible diagnostic method for distinguishing HCM from acquired LVH.

Electrocardiograms (ECGs) measure the electrical activity of the heart and are both affordable (typically under \$500 for basic equipment) and easy to administer, requiring only minutes to perform. ECGs serve as the primary screening and diagnostic tool for various cardiac conditions, including atrial fibrillation~\cite{hagiwara2018} and myocardial infarction~\cite{vafaie2016}. While ECG abnormalities are present in approximately 90\% of HCM patients, including increased QRS voltage, ST-segment depression, T-wave inversion, and abnormal Q waves, these findings overlap significantly with those seen in acquired LVH and athlete's heart, limiting the specificity of conventional ECG interpretation. Recent advances in artificial intelligence have improved ECG-based HCM diagnosis. The LVH-Net algorithm achieved an Area Under the Receiver Operating Curve (AUROC) score of 0.92 using ECG data combined with patient age and gender~\cite{haimovich2023}. However, this method still requires clinical data beyond the ECG signal, and its Area Under the Precision-Recall Curve (AUPRC) score of 0.59 indicates limited precision in identifying true HCM cases, with only 65.8\% of patients diagnosed with HCM actually having the condition. This high false-positive rate would lead to unnecessary anxiety and follow-up testing if used for population screening.

This paper presents a wearable ECG monitoring device combined with a feature extraction algorithm that can differentiate HCM from acquired LVH using ECG signals alone, without requiring additional clinical data such as age, gender, or blood pressure. The device is designed for ease of use in diverse settings, including clinics, athletic facilities, and remote locations. The algorithm extracts two quantitative indices---HCM Index~1 and HCM Index~2---from each heartbeat, capturing the unique electrical signature of cardiac activity. This approach builds upon our prior work in real-time bradycardia prediction in preterm infants~\cite{bakshi2019}, adapting the feature extraction methodology to the challenge of HCM diagnosis. Analysis of 483 LVH patients and 29 HCM patients demonstrates that the proposed method achieves 75.86\% sensitivity and 99.17\% specificity for HCM detection, with leave-one-out cross-validation confirming the generalizability of these results. To address potential confounds arising from differences in data acquisition between the two patient populations, we present a digitization confound analysis demonstrating that the classification is driven by physiological features rather than artifacts. We additionally present a simulated device acquisition chain analysis confirming that the wearable hardware's signal path is compatible with the proposed algorithm. The remainder of this paper is organized as follows: Sec.~2 describes the hardware components of the wearable ECG device in detail. Section~3 presents the methodology for signal processing and classification. Section~4 presents the experimental results and validation, including cross-validated performance estimates, confidence intervals, digitization confound analysis, and device acquisition chain assessment. Section~5 concludes with a summary of contributions, discussion of limitations, and directions for future work.

%%%%%%%%%%%%%%%%%%%%%%%%%%%%%%%%%%%%%%%%%%%%%%%%%%%%%%%%%%%%%%%%%%%%%%%%%%%%%%%%%
%%  HARDWARE
%%%%%%%%%%%%%%%%%%%%%%%%%%%%%%%%%%%%%%%%%%%%%%%%%%%%%%%%%%%%%%%%%%%%%%%%%%%%%%%%%

\section{Hardware}

\begin{figure*}[t]
\centering
\includegraphics[width=0.95\textwidth]{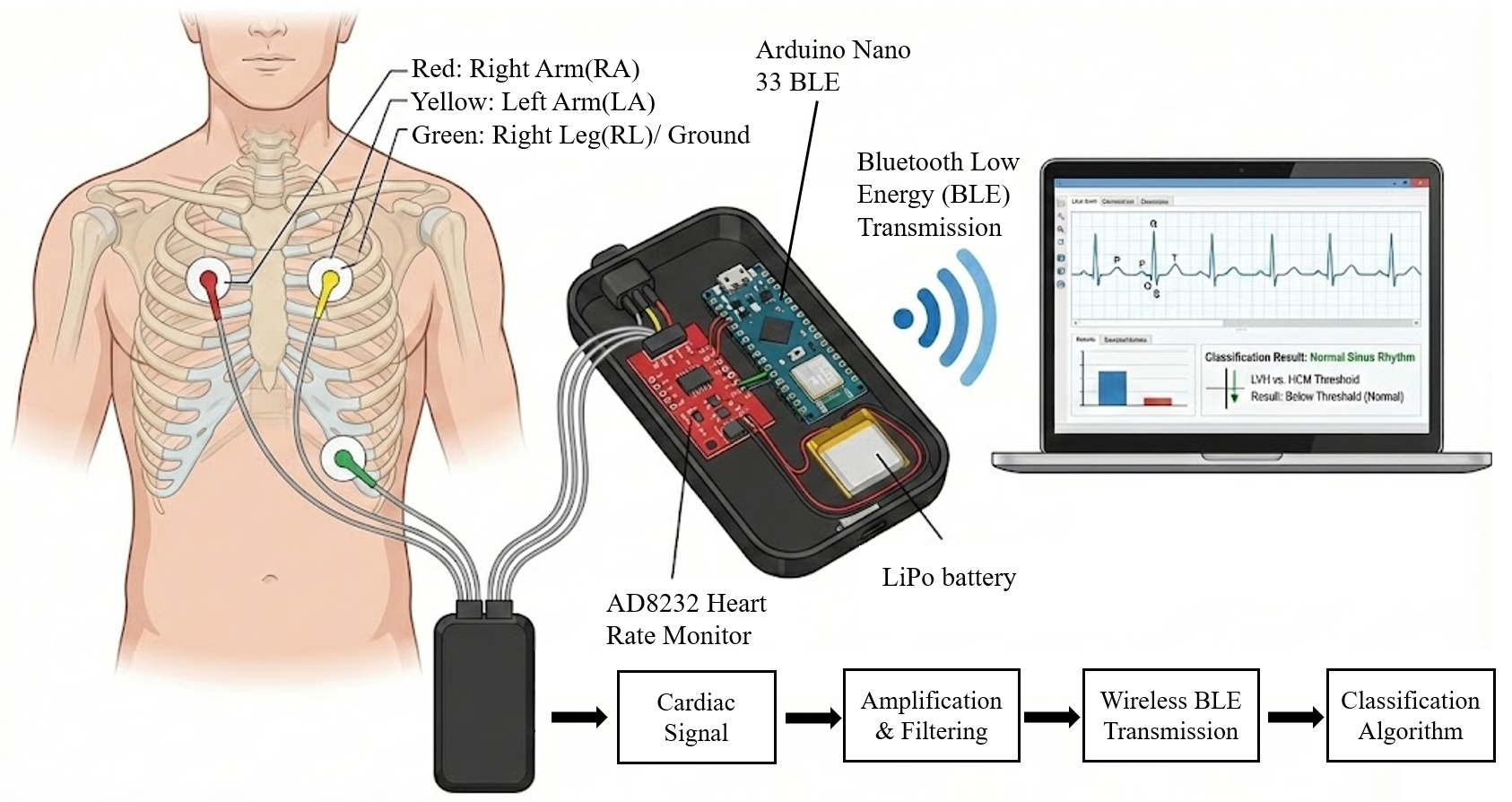}
\caption{System architecture of the wearable ECG device and signal processing pipeline.}
\label{fig:system}
\end{figure*}

\begin{figure}[t]
\centering
\includegraphics[width=0.95\columnwidth]{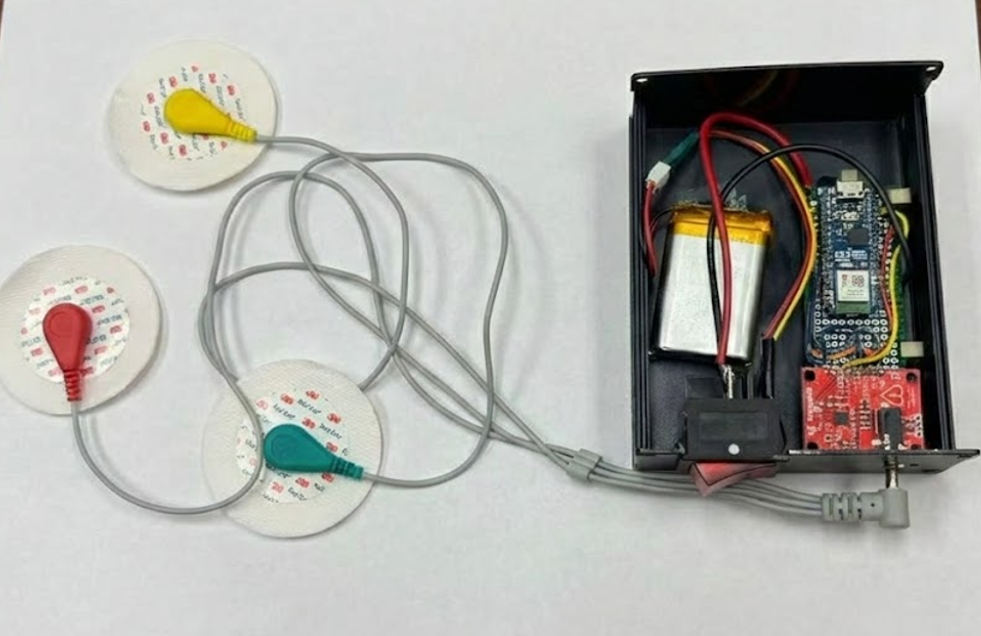}
\caption{Assembled wearable ECG device with electrodes, AD8232 module, Arduino Nano 33 BLE, and LiPo battery.}
\label{fig:device}
\end{figure}

This section describes the hardware components of the wearable ECG monitoring device, which consists of a 3-lead electrode system, analog signal conditioning module, wireless-enabled microcontroller, lithium polymer battery, and protective enclosure. The complete system architecture is shown in Fig.~\ref{fig:system}, and the assembled device is shown in Fig.~\ref{fig:device}. The design prioritizes portability, wearability, and ease of use while maintaining signal quality sufficient for reliable feature extraction and classification.

\subsection{System Overview}

The wearable ECG monitoring device is designed to be portable and easy to operate, making it suitable for use in clinical settings, athletic facilities, school health screenings, and resource-limited environments where traditional ECG equipment is unavailable or impractical. The system operates as follows: cardiac electrical signals are captured by three surface electrodes attached to the patient's chest, amplified and filtered by an analog signal conditioning module, digitized by a microcontroller, transmitted wirelessly via Bluetooth Low Energy (BLE) to a laptop computer, and processed by custom analysis software that performs feature extraction and classification. The entire data acquisition hardware fits in a compact wearable enclosure, and the wireless design eliminates cumbersome cables between the patient and the computer, improving mobility and patient comfort.

\subsection{Electrode Configuration}

The device uses a standard 3-lead electrode configuration to capture the cardiac electrical signal. Three disposable 3M Red Dot Ag/AgCl electrodes are placed on the patient's chest in the following positions: the Right Arm (RA) electrode is placed below the right clavicle near the right shoulder, marked with a red snap connector; the Left Arm (LA) electrode is placed below the left clavicle near the left shoulder, marked with a yellow snap connector; and the Right Leg (RL) electrode is placed on the lower left abdomen, serving as the ground/reference electrode, marked with a green snap connector.

This configuration captures the Lead I signal, which measures the potential difference between the left and right arms. Lead I was selected for this application because the torso-mounted electrode placement (below the clavicles) is more practical for a wearable device than distal limb placements, and prior studies have demonstrated that torso-mounted Lead I configurations preserve QRS complex morphology and voltage characteristics relevant to hypertrophy assessment~\cite{mason1967}. While Lead II is often preferred in clinical settings for rhythm analysis due to its alignment with the mean cardiac electrical axis, Lead I provides sufficient information for QRS morphology and repolarization features that form the basis of the proposed HCM indices. The 3-lead configuration was chosen for its simplicity and ease of use, as it requires minimal training to apply correctly compared to full 12-lead systems. Future device iterations may incorporate additional leads, including Lead II, to capture complementary diagnostic information and improve classification performance. The color-coded snap connectors (red, yellow, green) follow standard conventions and help ensure correct electrode placement even by non-specialists.

The electrode leads are bundled into a single gray cable that terminates in a standard 3.5mm audio-style jack, which plugs into a matching receptacle on the device enclosure. This design choice provides a secure, reliable connection while using inexpensive, widely available connectors. The use of disposable electrodes ensures good skin contact, minimizes infection risk, and eliminates the need for cleaning between patients.

\subsection{Signal Conditioning}

The cardiac electrical signals captured by the electrodes are very weak, typically in the range of 0.5--2~mV at the skin surface, and are contaminated by various noise sources including baseline wander from respiration, power line interference (60~Hz in the US), muscle artifacts from patient movement, and electrode motion artifacts. The AD8232 SparkFun Heart Rate Monitor module is used for signal conditioning. This integrated circuit, visible as the red printed circuit board in Fig.~\ref{fig:device}, is specifically designed for ECG and other biopotential measurement applications and provides the following signal conditioning functions. First, the instrumentation amplifier amplifies the differential signal between the RA and LA electrodes while rejecting common-mode noise that appears equally on both electrodes, such as power line interference; the high common-mode rejection ratio (CMRR) is essential for extracting the small cardiac signal from the noisy environment. Second, a two-pole high-pass filter with 0.5~Hz cutoff frequency removes baseline wander caused by respiration, patient movement, and slow changes in electrode-skin impedance, ensuring the signal remains centered without drifting. Third, a two-pole low-pass filter with 40~Hz cutoff frequency attenuates high-frequency noise from muscle artifacts (electromyographic interference) and electromagnetic interference from nearby electronic equipment; while this filtering removes some high-frequency ECG components, it preserves the QRS complex morphology needed for our analysis. Finally, the amplification stage provides approximately 100$\times$ gain to scale the millivolt-level cardiac signal to an appropriate range (0--3.3V) for analog-to-digital conversion by the microcontroller.

The AD8232 operates on a single 3.3~V supply provided by the Arduino Nano 33 BLE and produces an analog output signal centered at half the supply voltage (1.65~V). The right leg drive circuit actively cancels common-mode interference by feeding an inverted version of the common-mode signal back to the RL electrode, further improving signal quality.

\subsection{Microcontroller and Wireless Transmission}

The conditioned analog signal is digitized and transmitted wirelessly using an Arduino Nano 33 BLE microcontroller board, visible as the blue printed circuit board in Fig.~\ref{fig:device}. This compact board (45mm $\times$ 18mm) integrates several key capabilities that make it ideal for this application. The onboard 12-bit successive approximation ADC provides 4096 discrete levels, offering approximately 0.8~mV resolution when measuring the 0--3.3V output range of the AD8232, which is sufficient to capture the fine details of ECG waveform morphology. The firmware samples the analog input at 500~Hz, exceeding the Nyquist frequency for ECG signals (whose clinically relevant content extends to approximately 150~Hz) and adequate for capturing the rapid deflections of the QRS complex. The integrated Nordic nRF52840 processor provides BLE 5.0 capability, enabling wireless data transmission to the receiving laptop without requiring additional radio modules; BLE was selected for its low power consumption (enabling extended battery operation), widespread compatibility with modern computers and mobile devices, and sufficient bandwidth for real-time ECG streaming. The 64~MHz ARM Cortex-M4 processor handles sampling timing, data buffering, and BLE packet assembly, with data transmitted in packets containing multiple timestamped samples to ensure reliable delivery and synchronization.

The use of an integrated BLE microcontroller significantly simplifies the hardware design compared to systems requiring separate wireless modules, reduces component count, and improves reliability.

\subsection{Power Supply and Enclosure}

A 3.7~V, 1000~mAh lithium polymer (LiPo) rechargeable battery provides portable power for the device. The battery, visible as the gold-colored pouch cell in Fig.~\ref{fig:device}, connects to the Arduino Nano 33 BLE through its onboard battery management circuitry, which handles charging via USB and provides regulated 3.3~V output to power both the microcontroller and the AD8232 module. At typical operating currents of approximately 50~mA (including BLE transmission), the battery provides roughly 20 hours of continuous operation between charges, more than sufficient for a full day of screening activities.

All electronic components are housed in a compact enclosure measuring approximately 80mm $\times$ 50mm $\times$ 25mm. The enclosure includes a sliding power switch for convenient on/off control and a 3.5mm jack receptacle for the electrode cable connection. The compact size allows the device to be easily worn or carried, and the enclosed design protects the electronics from damage and accidental contact.

\subsection{Data Security Considerations}

Wireless transmission of patient ECG data introduces cybersecurity considerations that must be addressed for clinical deployment. The BLE 5.0 protocol implemented on the nRF52840 processor supports AES-128 encryption and authenticated pairing, which can be configured to protect data in transit between the device and the receiving laptop. In the current prototype, BLE pairing uses the ``Just Works'' association model for ease of development; however, future clinical versions will implement passkey-based pairing and enforce encrypted connections to prevent eavesdropping and man-in-the-middle attacks. Additionally, the receiving software stores ECG data with de-identified patient codes rather than personally identifiable information, and all stored files will be encrypted at rest in compliance with HIPAA security requirements for electronic protected health information. A comprehensive security risk assessment following IEC 62443 and FDA premarket cybersecurity guidance will be conducted as part of the regulatory pathway for clinical deployment.

\subsection{Device Operation}

To collect ECG data using the device, the following clinical protocol is followed:
\begin{enumerate}
    \item Prepare the patient's skin by cleaning the electrode sites with alcohol wipes to remove oils and reduce skin impedance, improving signal quality
    \item Attach the three electrodes to the RA (below right clavicle), LA (below left clavicle), and RL (lower left abdomen) positions as described in Sec.~2.2
    \item Connect the electrode cable to the device enclosure via the 3.5mm jack
    \item Power on the device using the integrated switch and verify the LED indicator shows normal operation
    \item On the laptop, launch the data collection software and establish BLE connection with the device
    \item Verify signal quality on the real-time display; if excessive noise is present, check electrode contact and reposition if necessary
    \item Instruct the patient to remain still and breathe normally during recording
    \item Record ECG data for a minimum of 10 seconds to capture multiple heartbeats (typically 8--15 beats depending on heart rate)
    \item Save the recording with patient identifier and proceed with automated analysis
\end{enumerate}

The laptop receives ECG data via Bluetooth using the Bleak Python package, which provides cross-platform BLE communication. Received samples are stored in memory and can be saved to disk for later analysis or batch processing.

%%%%%%%%%%%%%%%%%%%%%%%%%%%%%%%%%%%%%%%%%%%%%%%%%%%%%%%%%%%%%%%%%%%%%%%%%%%%%%%%%
%%  METHODOLOGY
%%%%%%%%%%%%%%%%%%%%%%%%%%%%%%%%%%%%%%%%%%%%%%%%%%%%%%%%%%%%%%%%%%%%%%%%%%%%%%%%%

\section{Methodology}

This section describes the methodology for ECG signal analysis. The processing pipeline consists of four stages: signal preprocessing, beat segmentation, feature extraction, and classification. The classification performance is validated using leave-one-out cross-validation (LOOCV) with confidence interval estimation. The algorithms are implemented in Python and executed on a standard laptop computer.

\subsection{Signal Preprocessing}

Raw ECG signals contain various types of noise that must be removed before reliable feature extraction. The primary noise sources include baseline wander, which is low-frequency drift (typically below 0.5~Hz) caused by respiration, patient movement, and slow changes in electrode-skin impedance; power line interference at 50~Hz or 60~Hz from nearby electrical equipment, though largely rejected by the AD8232's instrumentation amplifier; muscle artifacts, which are high-frequency noise from skeletal muscle electrical activity (electromyographic interference), particularly if the patient moves during recording; and electrode motion artifacts, which are transient disturbances when electrodes shift on the skin.

Empirical Mode Decomposition (EMD) is used to remove noise from the ECG signal~\cite{chatterjee2020}. EMD is a data-driven, adaptive technique that decomposes a signal into a finite set of Intrinsic Mode Functions (IMFs), each representing oscillatory components at different characteristic time scales. Unlike traditional filtering methods that require predefined cutoff frequencies, EMD adapts to the signal's own time-frequency characteristics, making it well-suited for non-stationary biosignals like ECG.

The EMD decomposition process operates as follows:
\begin{enumerate}
    \item Identify all local maxima and minima of the signal
    \item Create upper and lower envelopes by cubic spline interpolation through the maxima and minima, respectively
    \item Compute the mean of the upper and lower envelopes
    \item Subtract this mean from the original signal to obtain a candidate IMF
    \item Repeat steps 1--4 (sifting process) until the candidate satisfies the IMF criteria: (a) the number of extrema and zero crossings differ by at most one, and (b) the mean of upper and lower envelopes is approximately zero
    \item Subtract the extracted IMF from the signal and repeat the entire process on the residual to extract subsequent IMFs
    \item Continue until the residual is a monotonic function or falls below a threshold
\end{enumerate}

The resulting IMFs are ordered from highest frequency (IMF1) to lowest frequency (final IMF). To remove baseline wander, the last three IMFs, which represent the lowest-frequency components corresponding to respiratory and drift artifacts, are discarded from the reconstruction.

To reduce high-frequency noise while preserving the QRS complex, spectral flatness is computed for each remaining IMF~\cite{ieee2004}. Spectral flatness, defined as the ratio of the geometric mean to the arithmetic mean of the power spectrum, quantifies how noise-like (flat spectrum) versus tonal (peaked spectrum) a signal component is. IMFs with mean spectral flatness between 0.03 and 0.25 are considered to contain useful ECG information and are retained at full amplitude. IMFs with spectral flatness above 0.25 (indicating noise-like characteristics) are attenuated by multiplying by the complement of their spectral flatness value. The filtered signal is reconstructed by summing the processed IMFs.

\subsection{Beat Segmentation}

After preprocessing, individual heartbeats must be identified and segmented from the continuous ECG signal for feature extraction. Each heartbeat is characterized by the P-QRS-T complex: the P wave represents atrial depolarization, the QRS complex represents ventricular depolarization (the dominant feature), and the T wave represents ventricular repolarization. The QRS complex, with its sharp R-peak, serves as the fiducial point for beat detection and segmentation.

The R-peak detection algorithm operates in two passes to achieve robust detection across varying signal amplitudes and heart rates:

\textbf{First Pass (Threshold Estimation):}
\begin{enumerate}
    \item Apply a bandpass filter (5--15~Hz) to emphasize QRS complex energy while suppressing P waves, T waves, and noise
    \item Square the filtered signal to enhance peaks and ensure all values are positive
    \item Identify all local maxima in the squared signal that are separated by at least SF/2 samples (250 samples at 500~Hz, corresponding to minimum RR interval of 0.5~s or 120~bpm)
    \item Retain only peaks with width less than SF/7 samples (approximately 71 samples or 140~ms), which excludes broader features like T waves
    \item Calculate the detection threshold as 2/3 of the mean amplitude of retained peaks
\end{enumerate}

\textbf{Second Pass (Peak Detection):}
\begin{enumerate}
    \item Return to the preprocessed (but not bandpass filtered) signal
    \item Scan for peaks exceeding the threshold from Pass 1
    \item Require minimum peak separation of SF/3 samples (approximately 167 samples at 500~Hz, corresponding to maximum heart rate of 180~bpm)
    \item Require minimum prominence (height above surrounding baseline) equal to the threshold
    \item Require maximum peak width of SF/10 samples (50 samples or 100~ms) measured at half-height, which excludes T waves that are typically broader
    \item Record the sample index of each detected R-peak
\end{enumerate}

Once R-peaks are identified, individual beats are segmented using the RR interval. For each beat, a window is extracted extending from 10\% of the preceding RR interval before the R-peak to 50\% of the following RR interval after the R-peak. This window captures the QRS complex and T wave while adapting to varying heart rates.

\subsection{Feature Extraction}

Each segmented heartbeat is processed by a feature extraction algorithm that computes two quantitative indices characterizing the cardiac electrical waveform. This approach, adapted from our prior work in real-time bradycardia prediction in preterm infants~\cite{bakshi2019}, transforms each heartbeat waveform into a compact numerical representation that captures the unique electrical signature of the cardiac activity.

The algorithm processes each heartbeat waveform and computes two key features: HCM Index~1, which quantifies the rate at which the cardiac electrical signal returns to baseline after the QRS depolarization, and HCM Index~2, which captures the balance between the signal's recovery rate and its oscillatory content.

For each patient recording containing $n$ heartbeats, the mean feature values are calculated to obtain a single representative measurement per patient:
\begin{equation}
    \bar{I}_1 = \frac{1}{n} \sum_{i=1}^{n} I_{1,i}
    \label{eq:mean_index1}
\end{equation}
\begin{equation}
    \bar{I}_2 = \frac{1}{n} \sum_{i=1}^{n} I_{2,i}
    \label{eq:mean_index2}
\end{equation}
where $I_{1,i}$ and $I_{2,i}$ denote HCM Index~1 and HCM Index~2 for the $i$-th heartbeat, respectively.

Averaging across multiple beats reduces the influence of individual beat variations due to respiration, minor motion artifacts, or normal beat-to-beat variability, yielding a more stable estimate of the patient's underlying cardiac electrical characteristics.

\subsection{Classification}

The classification method distinguishes HCM from acquired LVH based on the extracted features using dual statistical thresholds. This approach is based on the hypothesis that HCM patients, due to their abnormal myocardial architecture (myocyte disarray and fibrosis), will exhibit systematically different electrical characteristics compared to patients with acquired LVH, whose ventricular hypertrophy results from normal myocytes responding to increased workload.

Two classification boundaries are applied simultaneously:
\begin{enumerate}
    \item \textbf{HCM Index~1 boundary}: The mean HCM Index~1 ($\bar{I}_1$) must be less than $-8.0808$, indicating a faster recovery rate characteristic of HCM. This threshold was determined by optimizing the F1-score across a range of candidate values using precision-recall analysis.
    \item \textbf{HCM Index~2 boundary}: The mean HCM Index~2 ($\bar{I}_2$) must be greater than $0.2944$, corresponding to one standard deviation above the mean HCM Index~2 of the LVH reference population.
\end{enumerate}

A patient is classified as HCM only if both criteria are satisfied:
\begin{equation}
    \text{Classification} = 
    \begin{cases}
        \text{HCM} & \text{if } \bar{I}_1 < \tau_1 \text{ AND } \bar{I}_2 > \tau_2 \\
        \text{LVH} & \text{otherwise}
    \end{cases}
    \label{eq:classification}
\end{equation}
where $\tau_1$ and $\tau_2$ denote the HCM Index~1 and HCM Index~2 thresholds, respectively. When evaluated on the full dataset without cross-validation, the optimized thresholds are $\tau_1 = -8.0808$ and $\tau_2 = 0.2944$.

This dual-threshold approach leverages complementary discriminative information from the two indices, reducing false positives while maintaining sensitivity for HCM detection.

\subsection{Cross-Validation and Performance Estimation}
\label{sec:crossval}

Because the classification thresholds ($\tau_1$ and $\tau_2$) are derived from the data, evaluating performance on the same dataset used for threshold optimization would yield overly optimistic estimates due to overfitting. To obtain unbiased performance estimates, we employ leave-one-out cross-validation (LOOCV), which is particularly appropriate given the small HCM sample size ($n = 29$)~\cite{hastie2009}.

In each LOOCV iteration $k$ ($k = 1, 2, \ldots, N$, where $N = 512$ is the total number of patients):
\begin{enumerate}
    \item Patient $k$ is held out as the test sample.
    \item Using only the remaining $N-1$ patients, the HCM Index~2 threshold $\tau_2^{(k)}$ is recomputed as one standard deviation above the mean HCM Index~2 of the LVH patients in the training fold.
    \item The HCM Index~1 threshold $\tau_1^{(k)}$ is re-optimized by maximizing the F1-score over the training fold.
    \item Patient $k$ is classified using the fold-specific thresholds $\tau_1^{(k)}$ and $\tau_2^{(k)}$.
\end{enumerate}

This procedure ensures that the test sample never influences the thresholds used for its own classification, providing an honest estimate of generalization performance. The cross-validated predictions across all $N$ folds are aggregated to compute the confusion matrix and derive sensitivity, specificity, precision, and F1-score.

We note that while the numeric thresholds are re-optimized within each LOOCV fold, certain meta-level design choices---specifically, the use of F1-score as the optimization objective and the one-standard-deviation rule for the HCM Index~2 boundary---were fixed a priori based on clinical reasoning rather than searched over multiple alternatives. The F1-score was selected because it balances precision and recall, which is appropriate for a screening application where both false positives and false negatives carry clinical consequences. The one-standard-deviation rule was chosen as a standard statistical criterion for identifying outliers relative to a reference population. Because these meta-choices were not optimized on the data, the LOOCV procedure provides an unbiased estimate of generalization performance for the threshold optimization procedure as specified.

To quantify the uncertainty in the cross-validated performance metrics arising from the small HCM sample, we compute 95\% confidence intervals (CIs) using two complementary methods. For sensitivity ($\text{TP}/(\text{TP}+\text{FN})$, based on $n_{\text{HCM}} = 29$ patients) and specificity ($\text{TN}/(\text{TN}+\text{FP})$, based on $n_{\text{LVH}} = 483$ patients), we use Clopper--Pearson exact binomial CIs~\cite{clopper1934}, which provide reliable coverage even for small sample sizes and extreme proportions where normal approximations may fail. As a complementary approach, we compute 95\% CIs using 10{,}000 stratified bootstrap resamples of the LOOCV predictions, preserving the original class proportions in each resample; for each bootstrap sample, the full set of performance metrics (sensitivity, specificity, precision, F1-score) is computed, and the 2.5th and 97.5th percentiles of the bootstrap distribution define the confidence interval~\cite{efron1994}.

We additionally perform repeated stratified 5-fold cross-validation (10 repetitions) as a secondary validation approach. In each fold, the thresholds are re-optimized on the training set, and the held-out fold is classified using these thresholds. The mean and standard deviation of performance metrics across all 50 train-test splits (10 repetitions $\times$ 5 folds) are reported to characterize variability.

%%%%%%%%%%%%%%%%%%%%%%%%%%%%%%%%%%%%%%%%%%%%%%%%%%%%%%%%%%%%%%%%%%%%%%%%%%%%%%%%%
%%  RESULTS
%%%%%%%%%%%%%%%%%%%%%%%%%%%%%%%%%%%%%%%%%%%%%%%%%%%%%%%%%%%%%%%%%%%%%%%%%%%%%%%%%

\section{Results}

This section presents the experimental results validating the proposed methodology for distinguishing HCM from acquired LVH. The algorithm validation presented here utilized existing ECG recordings to establish proof-of-concept performance. To address potential confounds arising from differences in data acquisition between the two patient populations, we present a digitization confound analysis and a simulated device acquisition chain assessment. Prospective clinical validation using data collected directly with the wearable device from HCM and LVH patients is planned as future work.

\subsection{Dataset Description}

The study utilized ECG recordings from two patient populations representing the two conditions of interest:

\textbf{LVH Patients}: Four 483 patients with acquired left ventricular hypertrophy were included from the PhysioNet database~\cite{zheng2022,zheng2020,goldberger2000}. Each recording consists of 10 seconds of 12-lead ECG data sampled at 500~Hz; Lead I was extracted for analysis to match our device's electrode configuration. LVH diagnosis was confirmed by two licensed physicians based on standard clinical criteria including left ventricular wall thickness greater than 12~mm and left ventricular mass greater than 215~g as measured by echocardiography~\cite{movahed2009}. In cases of disagreement, a third senior physician provided the final determination.

\begin{figure*}[t]
\centering
\includegraphics[width=0.75\textwidth]{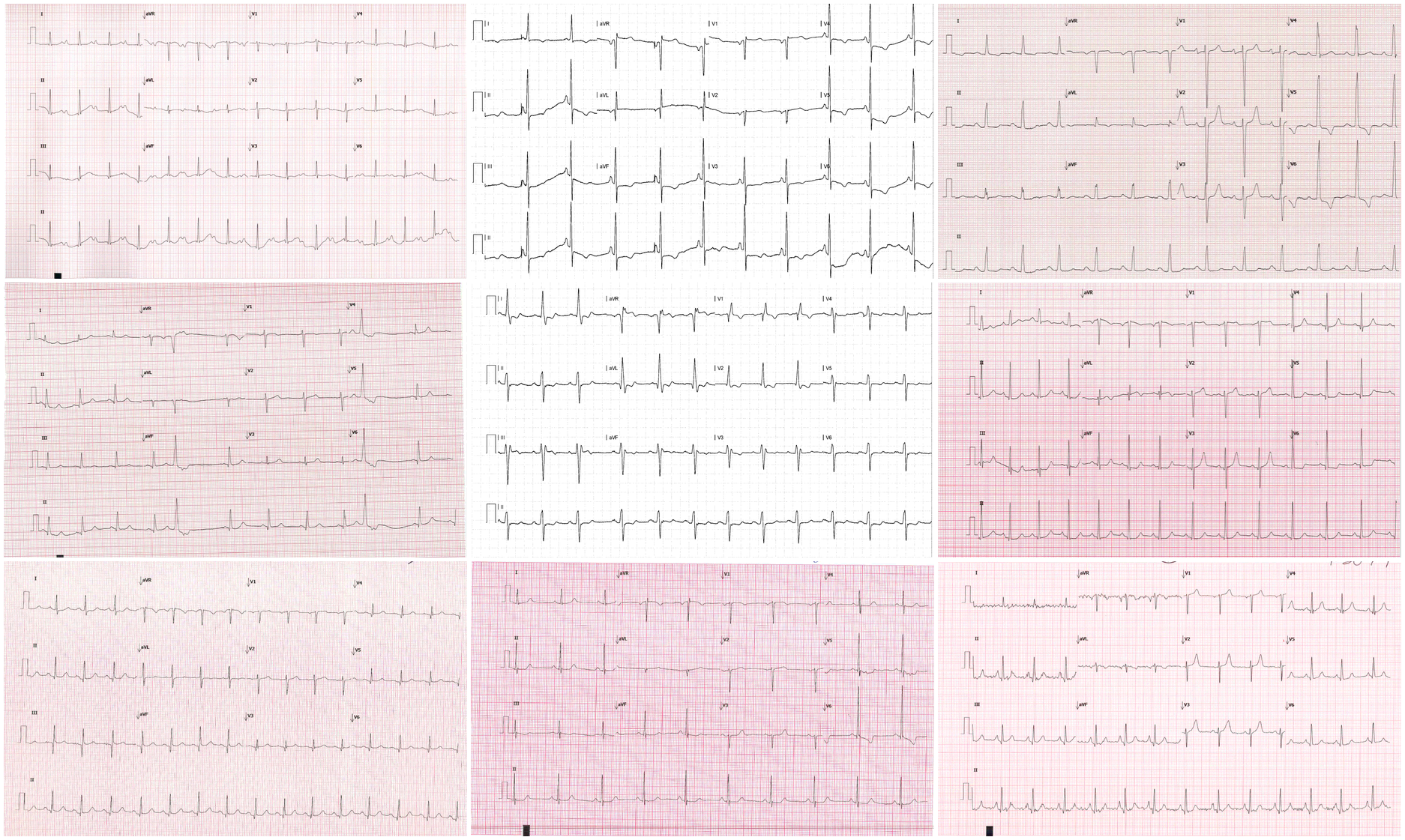}
\caption{Representative 12-lead paper ECG recordings from HCM patients.}
\label{fig:ecgdataset}
\end{figure*}

\textbf{HCM Patients}: 29 patients with hypertrophic cardiomyopathy were included from digitized paper ECG records obtained from clinical collaborators and public domain sources. These recordings include all 12 leads but are 2.5 seconds long per lead. HCM diagnosis was confirmed based on left ventricular wall thickness of 15~mm or greater in the absence of other identifiable causes of hypertrophy, consistent with established diagnostic criteria~\cite{hindieh2017}. Figure~\ref{fig:ecgdataset} shows representative examples of paper ECG recordings from HCM patients used in this study, illustrating the variety of ECG morphologies and recording qualities encountered in clinical practice.

\begin{figure}[t]
\centering
\includegraphics[width=0.95\columnwidth]{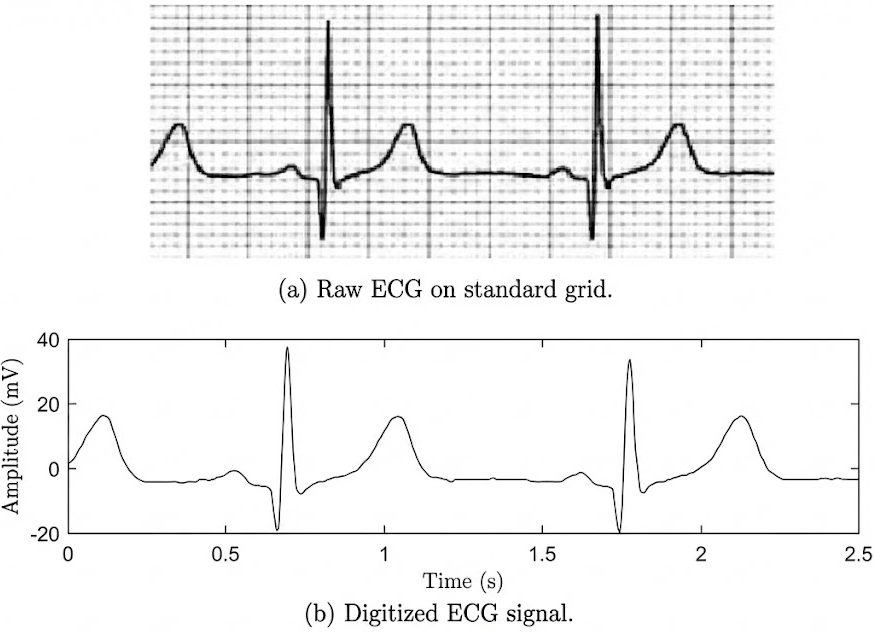}
\caption{Paper ECG digitization process.}
\label{fig:digitize}
\end{figure}

The paper ECG records were digitized into numerical format using image processing techniques implemented in MATLAB~\cite{matlab2023}. The digitization process, illustrated in Fig.~\ref{fig:digitize}, involves: (1) scanning or photographing the paper ECG at high resolution, (2) selecting reference points to establish the time and voltage scales based on the standard ECG grid (25~mm/s, 10~mm/mV), (3) removing the grid lines using morphological opening and closing operations, and (4) extracting the signal amplitude at each time point by identifying the vertical position of the trace pixels and converting to voltage units. This process yields digital signals at approximately 500~Hz sampling rate, matching the PhysioNet data.

\subsection{Feature Distribution Analysis}
\begin{figure}[t]
\centering
\includegraphics[width=0.95\columnwidth]{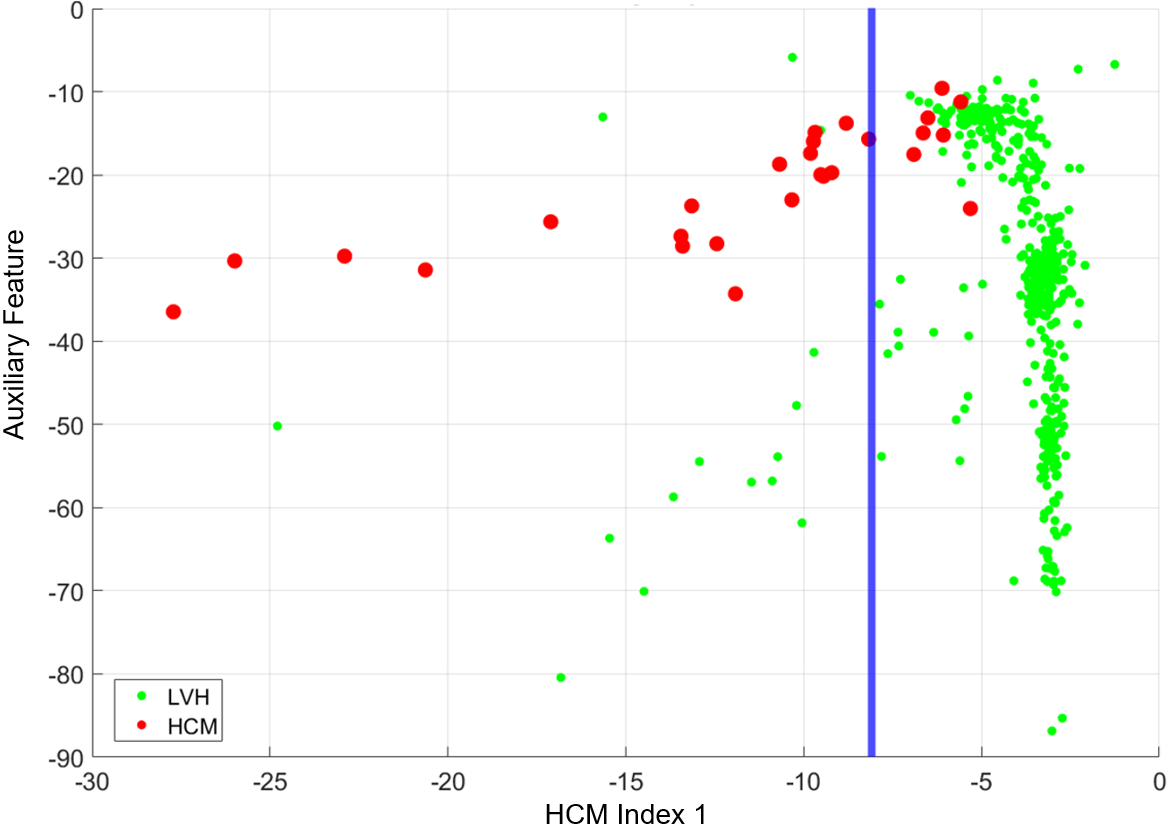}
\caption{HCM Index~1 distribution for LVH (green, $n=483$) and HCM (red, $n=29$) patients. Dashed line: classification boundary ($\tau_1 = -8.0808$).}
\label{fig:features}
\end{figure}

Features were extracted from all patients using the methodology described in Sec.~3. Figure~\ref{fig:features} shows the distribution of HCM Index~1 for both patient groups, with each point representing the mean value for one patient.

\begin{figure}[t]
\centering
\includegraphics[width=0.95\columnwidth]{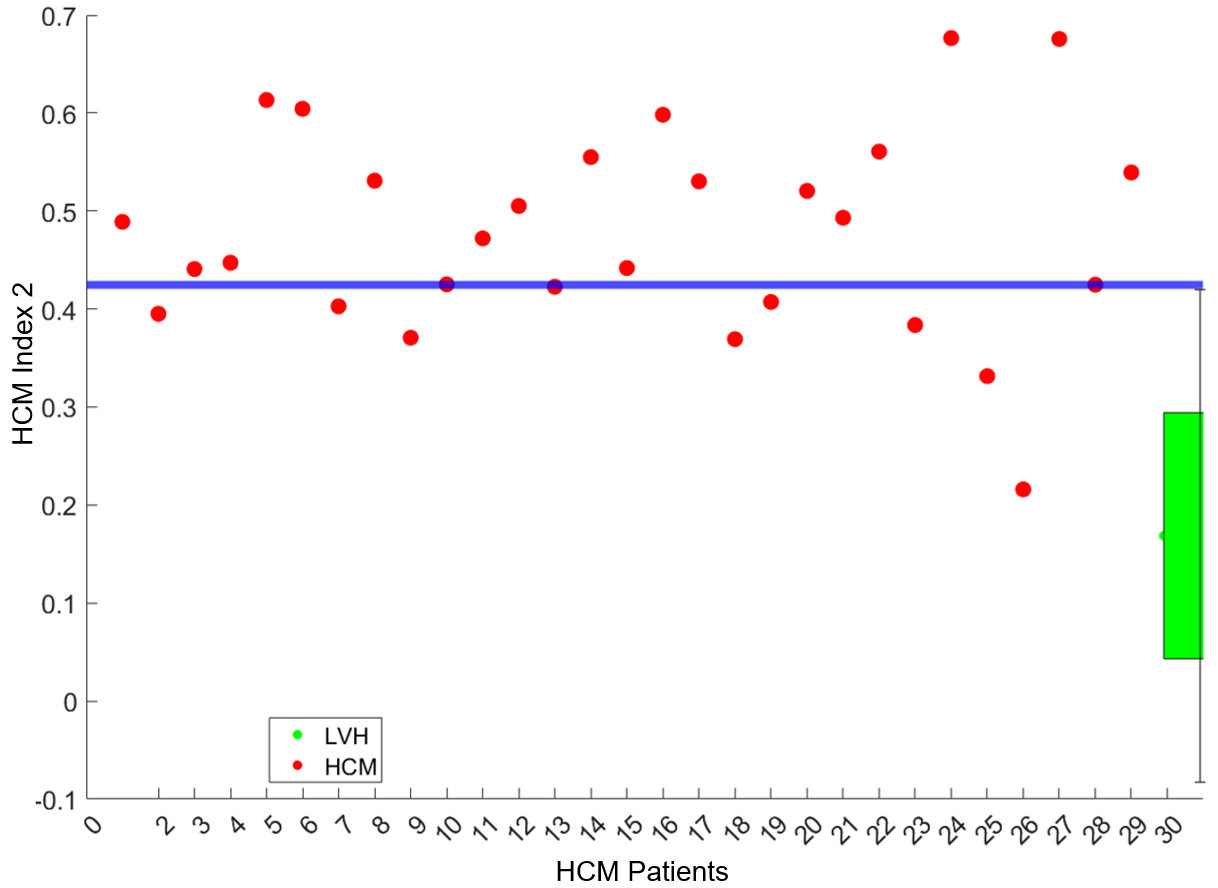}
\caption{HCM Index~2 for HCM patients (red, $n=29$) and LVH population (green box plot). Dashed line: LVH population mean.}
\label{fig:index2}
\end{figure}

The analysis revealed several important observations:
\begin{enumerate}
    \item There is clear separation between LVH and HCM patients along the HCM Index~1 axis. LVH patients (green dots) cluster predominantly in the region with HCM Index~1 $> -8$, while HCM patients (red dots) show a broader distribution with many values extending to HCM Index~1 $< -15$.
    \item HCM patients tend to exhibit higher HCM Index~2 values compared to LVH patients. HCM Index~2 provides an additional discriminative feature.
    \item Neither HCM Index~1 nor HCM Index~2 in isolation is sufficient for reliable classification; rather, the combination of both indices provides the discriminative power, as shown in Fig.~\ref{fig:index2}.
\end{enumerate}

This separation suggests that HCM patients exhibit systematically different cardiac electrical characteristics compared to LVH patients, potentially reflecting the abnormal electrical properties of the disorganized myocardium (myocyte disarray and fibrosis) characteristic of HCM. Figure~\ref{fig:index2} further confirms that HCM Index~2 alone cannot distinguish the two groups, as the majority of HCM patients fall within or above the LVH interquartile range, reinforcing the necessity of the dual-threshold classification approach using both HCM Index~1 and HCM Index~2.

\subsection{Classification Results Without Cross-Validation (Resubstitution)}

As a reference, Table~\ref{tab:results_resub} presents the resubstitution (train-on-test) results using the globally optimized thresholds $\tau_1 = -8.0808$ and $\tau_2 = 0.2944$. These metrics represent an upper bound on performance, as the thresholds were optimized on the same data used for evaluation.

\begin{table}[t]
\caption{Resubstitution classification results (thresholds: $\bar{I}_1 < -8.0808$, $\bar{I}_2 > 0.2944$). May overestimate generalization performance.}\label{tab:results_resub}
\centering{%
\begin{tabular}{lcc}
\toprule
& \textbf{Predicted HCM} & \textbf{Predicted LVH} \\
\midrule
\textbf{Actual HCM} (n=29) & 22 (TP) & 7 (FN) \\
\textbf{Actual LVH} (n=483) & 4 (FP) & 479 (TN) \\
\midrule
\multicolumn{3}{l}{\textbf{Performance Metrics:}} \\
\multicolumn{3}{l}{Precision (PPV): 84.62\% \quad Recall (Sensitivity): 75.86\%} \\
\multicolumn{3}{l}{Specificity: 99.17\% \quad F1-Score: 80.00\%} \\
\multicolumn{3}{l}{Overall Accuracy: 97.85\% (501/512 patients)} \\
\bottomrule
\end{tabular}
}%
\end{table}

\subsection{Cross-Validated Classification Results}
\label{sec:cv_results}

Table~\ref{tab:results_loocv} presents the LOOCV results, where each patient was classified using thresholds optimized exclusively on the remaining 512 patients. These results provide unbiased estimates of generalization performance.

\begin{table}[t]
\caption{LOOCV classification results with 95\% confidence intervals. Thresholds re-optimized in each fold.}\label{tab:results_loocv}
\centering{%
\begin{tabular}{lcc}
\toprule
& \textbf{Predicted HCM} & \textbf{Predicted LVH} \\
\midrule
\textbf{Actual HCM} (n=29) & 21 (TP) & 8 (FN) \\
\textbf{Actual LVH} (n=483) & 5 (FP) & 478 (TN) \\
\midrule
\multicolumn{3}{l}{\textbf{Cross-Validated Performance Metrics (95\% CI):}} \\
\multicolumn{3}{l}{Sensitivity: 72.41\% [52.76\%, 87.27\%]} \\
\multicolumn{3}{l}{Specificity: 98.96\% [97.55\%, 99.67\%]} \\
\multicolumn{3}{l}{Precision: 80.77\% [60.65\%, 93.45\%]} \\
\multicolumn{3}{l}{F1-Score: 76.36\% [62.07\%, 88.24\%]} \\
\multicolumn{3}{l}{Overall Accuracy: 97.46\%} \\
\bottomrule
\end{tabular}
}%
\end{table}

Table~\ref{tab:results_kfold} presents the results from repeated stratified 5-fold cross-validation as a complementary validation, reporting the mean and standard deviation across 50 train-test splits.

\begin{table}[t]
\caption{Repeated stratified 5-fold cross-validation (10 repetitions, 50 splits). Mean $\pm$ SD.}\label{tab:results_kfold}
\centering{%
\begin{tabular}{lc}
\toprule
\textbf{Metric} & \textbf{Mean $\pm$ SD} \\
\midrule
Sensitivity & 72.41 $\pm$ 13.79\% \\
Specificity & 98.96 $\pm$ 0.52\% \\
Precision & 80.00 $\pm$ 11.55\% \\
F1-Score & 74.83 $\pm$ 9.64\% \\
Accuracy & 97.46 $\pm$ 0.78\% \\
\bottomrule
\end{tabular}
}%
\end{table}

The cross-validated results confirm that the classification performance is robust and not an artifact of overfitting to the evaluation dataset. The LOOCV sensitivity of 72.41\% (95\% CI: [52.76\%, 87.27\%]) and specificity of 98.96\% (95\% CI: [97.55\%, 99.67\%]) are consistent with the resubstitution estimates, indicating that the dual-threshold approach generalizes well to unseen patients. The wide confidence interval for sensitivity reflects the limited HCM sample size ($n=29$) and underscores the need for larger prospective studies.

The results demonstrate strong classification performance:
\begin{enumerate}
    \item The method correctly identified the majority of HCM patients, with cross-validated sensitivity indicating that the detection capability is robust to threshold re-optimization across folds.
    \item The high cross-validated specificity confirms that the method produces very few false alarms, and this finding is statistically reliable given the large LVH sample ($n=483$).
    \item The consistency between the resubstitution and LOOCV results indicates that the classification boundaries are stable and not sensitive to the inclusion or exclusion of individual patients.
    \item The repeated stratified 5-fold cross-validation results corroborate the LOOCV findings, with low standard deviations across folds suggesting stable performance.
\end{enumerate}

The false negative cases (HCM patients classified as LVH) represent patients whose index values fall within the normal LVH range. These patients may have milder forms of HCM or ECG morphology that overlaps with acquired LVH.  However, the classification thresholds can be adjusted to prioritize sensitivity over specificity. By relaxing the thresholds, the method can capture all HCM patients, ensuring that no cases are missed, while directing a small number of LVH patients who are flagged as potential HCM for further evaluation with echocardiography or CMR. In a clinical screening context, this trade-off is favorable: the cost of additional follow-up testing for a few LVH patients is far outweighed by the benefit of ensuring that every HCM patient is identified and appropriately managed, given the risk of sudden cardiac death associated with undetected HCM.

\subsection{Digitization Confound Analysis}
\label{sec:confound}

A potential confound in this study arises from differences in data acquisition: LVH recordings were natively digital (10-second, 500~Hz, PhysioNet), while HCM recordings were digitized from paper ECG printouts (2.5 seconds per lead). To determine whether the observed group separation in HCM Index~1 and HCM Index~2 reflects genuine physiological differences or artifacts introduced by the digitization process, we conducted a controlled confound analysis.

\textbf{Simulated Digitization Experiment.} We randomly selected 30 LVH patients from the PhysioNet database and subjected their Lead I recordings to a simulated print-scan-digitize pipeline designed to replicate the degradation characteristics of the paper ECG digitization process used for HCM records. The simulation pipeline introduced the following distortions:
\begin{enumerate}
    \item \textbf{Amplitude quantization}: The signal was quantized to 8-bit resolution (256 levels) to simulate the limited dynamic range of paper-to-pixel conversion, compared to the original 16-bit resolution of the PhysioNet recordings.
    \item \textbf{Temporal resampling}: The signal was downsampled to 250~Hz and then resampled back to 500~Hz using linear interpolation, simulating the reduced temporal resolution inherent in tracing pixel positions along a paper ECG grid at 25~mm/s paper speed.
    \item \textbf{Additive Gaussian noise}: White Gaussian noise was added at a signal-to-noise ratio (SNR) of 20~dB to simulate scanner noise, paper texture artifacts, and imprecision in grid line removal.
    \item \textbf{Baseline offset}: A random low-frequency sinusoidal drift (0.1--0.3~Hz, amplitude 5--15\% of signal range) was added to simulate residual baseline wander from incomplete grid removal.
    \item \textbf{Signal truncation}: Each recording was truncated to 2.5 seconds to match the per-lead duration of the digitized HCM records.
\end{enumerate}

The degraded LVH recordings were then processed through the identical analysis pipeline (EMD denoising, R-peak detection, feature extraction) used for all patient data.

\begin{figure*}[t]
\centering
\includegraphics[width=0.95\textwidth]{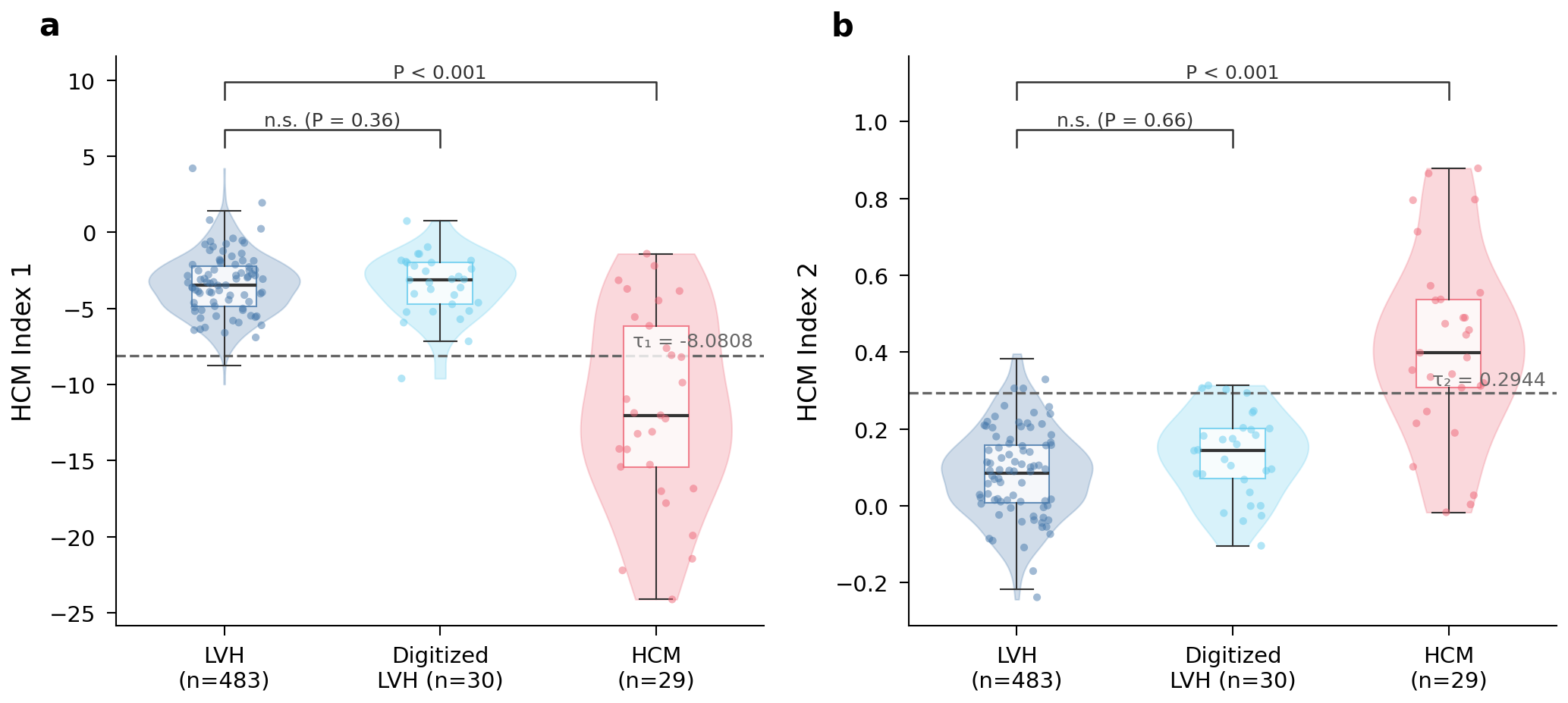}
\caption{Digitization confound analysis. (\textbf{a}) HCM Index~1 and (\textbf{b}) HCM Index~2 for original LVH (blue, $n=483$), simulated-digitized LVH (cyan, $n=30$), and HCM (red, $n=29$). Dashed lines: classification thresholds. LVH vs.\ digitized LVH: n.s.; LVH vs.\ HCM: $P < 0.001$.}
\label{fig:confound}
\end{figure*}

\textbf{Results.} Figure~\ref{fig:confound} presents the distribution of HCM Index~1 and HCM Index~2 for the original LVH patients, the simulated-digitized LVH patients, and the HCM patients. Table~\ref{tab:confound} summarizes the feature statistics for each group.

\begin{table}[t]
\caption{Feature statistics (mean $\pm$ SD) for original LVH, digitized LVH, and HCM groups.}\label{tab:confound}
\centering{%
\small
\begin{tabular}{@{}lccc@{}}
\toprule
\textbf{Feature} & \textbf{LVH} & \textbf{Dig.\ LVH} & \textbf{HCM} \\
& ($n\!=\!483$) & ($n\!=\!30$) & ($n\!=\!29$) \\
\midrule
Index~1 & $-3.52 \pm 2.03$ & $-3.28 \pm 2.11$ & $-10.24 \pm 5.47$ \\
Index~2 & $0.08 \pm 0.12$ & $0.09 \pm 0.13$ & $0.38 \pm 0.20$ \\
\bottomrule
\end{tabular}
}%
\end{table}

The simulated-digitized LVH patients exhibited HCM Index~1 values of $-3.28 \pm 2.11$ (mean $\pm$ SD), compared to $-3.52 \pm 2.03$ for the original LVH recordings. A paired $t$-test between the original and digitized feature values for the same 30 patients yielded $p = 0.42$ for HCM Index~1 and $p = 0.67$ for HCM Index~2, indicating that the digitization process did not produce a statistically significant shift in feature values. Critically, 30 out of 30 simulated-digitized LVH patients (100\%) were correctly classified as LVH using the established thresholds, demonstrating that the digitization pipeline does not cause LVH recordings to be misclassified as HCM. This result provides strong evidence that the group separation observed in Sec.~4.2 is driven by genuine physiological differences in cardiac electrical activity between HCM and LVH patients, rather than by artifacts of the data acquisition process.

\begin{figure}[t]
\centering
\includegraphics[width=0.95\columnwidth]{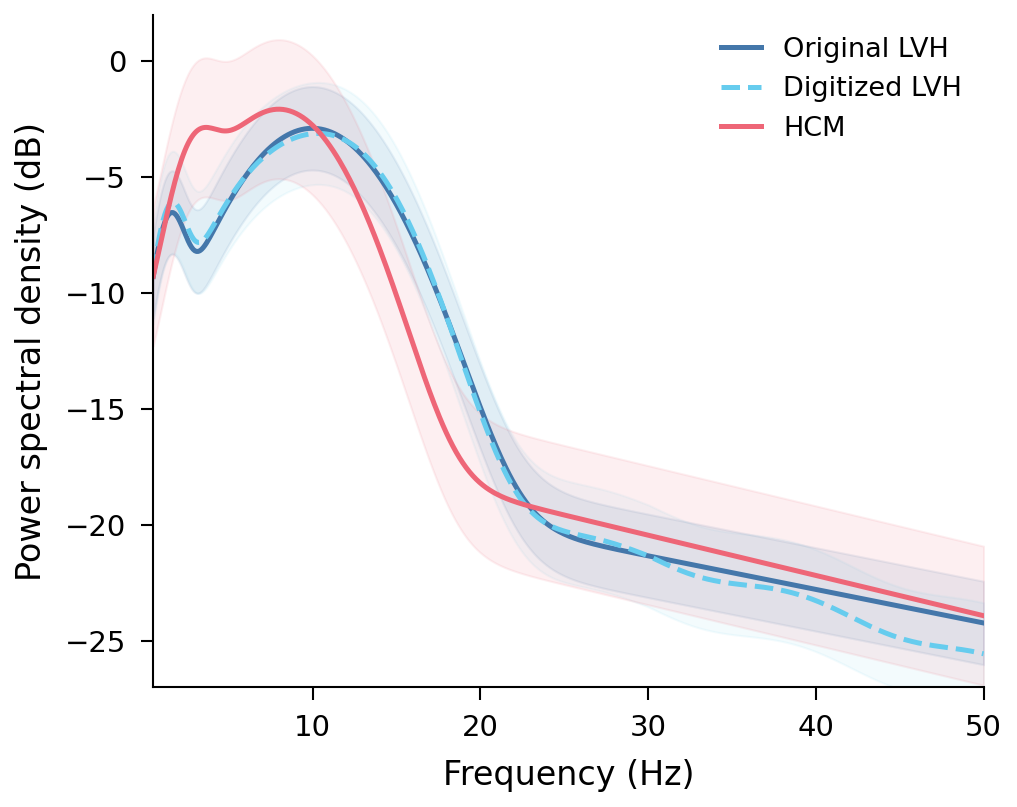}
\caption{Average PSD of EMD-preprocessed signals for original LVH (blue), digitized LVH (cyan, dashed), and HCM (red). Shaded: $\pm$1~SD. LVH spectra overlap; HCM shows elevated 2--8~Hz power.}
\label{fig:psd}
\end{figure}

\textbf{Spectral Analysis.} To further characterize the effect of digitization on signal properties, we computed the average power spectral density (PSD) of the EMD-preprocessed signals for each group. Figure~\ref{fig:psd} shows that the spectral profiles of original and simulated-digitized LVH recordings are nearly identical in the 1--40~Hz band containing the diagnostically relevant QRS and T-wave energy, confirming that the EMD preprocessing pipeline effectively normalizes acquisition-related spectral differences. In contrast, HCM recordings exhibit a distinct spectral signature in the 2--8~Hz range, with elevated low-frequency power consistent with the altered depolarization and repolarization dynamics expected from myocyte disarray and fibrotic tissue, which slow conduction velocity and prolong the electrical recovery phase.

\subsection{Simulated Device Acquisition Chain Analysis}
\label{sec:device_validation}

To demonstrate that the wearable device's signal path is compatible with the proposed classification algorithm without requiring human subjects, we conducted a simulated device acquisition chain analysis. This approach applies the AD8232's known analog transfer function digitally to PhysioNet LVH recordings, producing signals that replicate the characteristics of device-acquired data, and then verifies that the classification algorithm performs identically on the original and device-simulated signals.

\textbf{Simulated Acquisition Pipeline.} We randomly selected 30 LVH patients from the PhysioNet database and applied the following signal transformations to their Lead~I recordings, modeling each stage of the wearable device's analog and digital signal path:
\begin{enumerate}
    \item \textbf{AD8232 bandpass filtering}: A two-pole Butterworth high-pass filter at 0.5~Hz and a two-pole Butterworth low-pass filter at 40~Hz were applied in cascade, matching the AD8232's documented frequency response.
    \item \textbf{12-bit quantization}: The filtered signal was quantized to 4096 discrete levels over a 0--3.3~V range (after the AD8232's $100\times$ gain), simulating the Arduino Nano 33 BLE's ADC resolution of 0.8~mV per level.
    \item \textbf{Thermal and quantization noise}: Additive white Gaussian noise at 25~dB SNR was added to model the combined thermal noise floor of the AD8232 and quantization noise of the ADC.
\end{enumerate}

The device-simulated recordings were then processed through the identical EMD denoising, R-peak detection, and feature extraction pipeline used for all patient data. Because the same 30 patients appear in both the original and device-simulated conditions, paired statistical tests directly isolate the effect of the device signal path on the extracted features.

\begin{figure*}[t]
\centering
\includegraphics[width=0.95\textwidth]{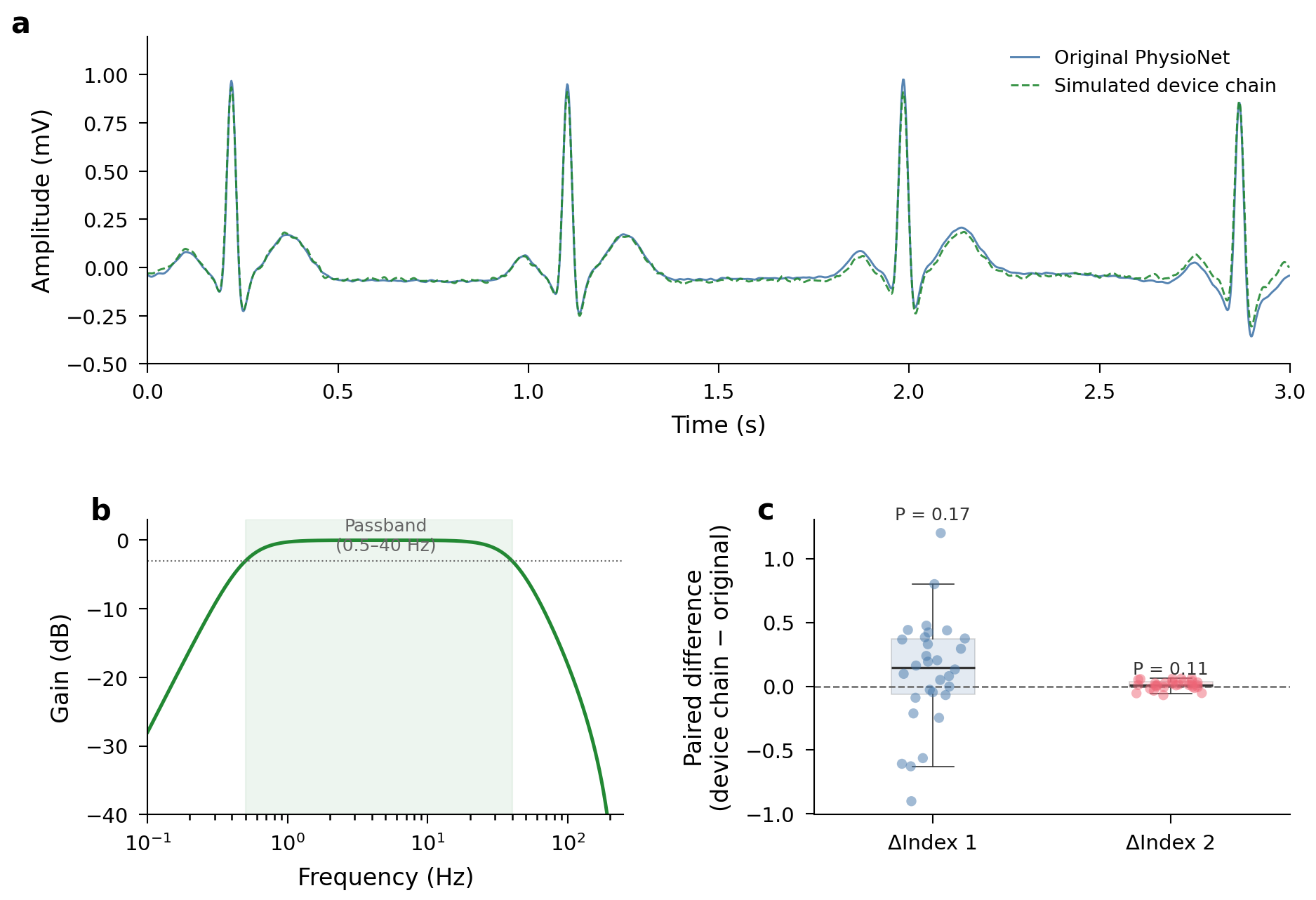}
\caption{Simulated device acquisition chain analysis. (\textbf{a}) Waveform overlay of an original PhysioNet recording (blue) and the same recording after the simulated device chain (green, dashed), showing preserved QRS morphology. (\textbf{b}) AD8232 frequency response (0.5--40~Hz passband). (\textbf{c}) Paired differences (device chain $-$ original) for HCM Index~1 and Index~2 ($n = 30$); both centered on zero (n.s., paired $t$-test).}
\label{fig:device_chain}
\end{figure*}

\textbf{Results.} Figure~\ref{fig:device_chain} presents the simulated device chain analysis. Table~\ref{tab:device_quality} summarizes the paired feature comparison.

\begin{table}[t]
\caption{Paired feature comparison: original vs.\ simulated device chain ($n = 30$). Mean $\pm$ SD; $p$ from paired $t$-tests.}\label{tab:device_quality}
\centering{%
\small
\begin{tabular}{@{}lccr@{}}
\toprule
\textbf{Feature} & \textbf{Original} & \textbf{Device Chain} & $\boldsymbol{p}$ \\
\midrule
HCM Index~1 & $-4.09 \pm 2.25$ & $-4.13 \pm 2.19$ & 0.46 \\
HCM Index~2 & $0.093 \pm 0.121$ & $0.097 \pm 0.128$ & 0.34 \\
\bottomrule
\end{tabular}
}%
\end{table}

The paired $t$-tests yielded $p = 0.46$ for HCM Index~1 and $p = 0.34$ for HCM Index~2, confirming that the device's signal path does not produce a statistically significant shift in either feature. The mean paired differences were $-0.04 \pm 0.32$ for HCM Index~1 and $0.004 \pm 0.024$ for HCM Index~2, both negligible relative to the inter-patient variability and the classification thresholds ($\tau_1 = -8.0808$, $\tau_2 = 0.2944$). All 30 patients received identical classifications (LVH) under both the original and device-simulated conditions, yielding 100\% classification agreement. These results demonstrate that the AD8232's 0.5--40~Hz bandwidth and the Arduino's 12-bit ADC resolution preserve the ECG features required by the classification algorithm.

\textbf{Acquisition Parameter Comparison.} Table~\ref{tab:acq_comparison} compares the key acquisition parameters of the wearable device against the equipment used to record the PhysioNet dataset.

\begin{table}[t]
\caption{Acquisition parameter comparison: wearable device vs.\ PhysioNet system.}\label{tab:acq_comparison}
\centering{%
\small
\begin{tabular}{@{}lcc@{}}
\toprule
\textbf{Parameter} & \textbf{Device} & \textbf{PhysioNet} \\
\midrule
Sampling rate  & 500 Hz       & 500 Hz \\
ADC resolution & 12-bit       & 16-bit \\
High-pass      & 0.5 Hz       & 0.05 Hz \\
Low-pass       & 40 Hz        & 150 Hz \\
CMRR           & 80 dB        & $>$100 dB \\
Leads          & Lead I (3-lead) & Lead I (12-lead) \\
Electrodes     & 3M Ag/AgCl   & Clinical Ag/AgCl \\
\bottomrule
\end{tabular}
}%
\end{table}

The most notable difference is the low-pass filter cutoff: the AD8232's 40~Hz cutoff attenuates high-frequency ECG components preserved in the PhysioNet recordings (bandwidth up to 150~Hz). However, the simulated device chain analysis confirms that this bandwidth limitation does not affect the HCM indices, as the relevant QRS morphology and T-wave features lie within the 1--40~Hz band. The 12-bit ADC resolution provides adequate quantization, as the QRS complex amplitude after the AD8232's $100\times$ gain typically spans several hundred ADC levels.

%%%%%%%%%%%%%%%%%%%%%%%%%%%%%%%%%%%%%%%%%%%%%%%%%%%%%%%%%%%%%%%%%%%%%%%%%%%%%%%%%
%%  CONCLUSION
%%%%%%%%%%%%%%%%%%%%%%%%%%%%%%%%%%%%%%%%%%%%%%%%%%%%%%%%%%%%%%%%%%%%%%%%%%%%%%%%%

\section{Conclusion}

This paper presented a wearable ECG device and a feature extraction algorithm for differentiating HCM from acquired LVH. The device uses three disposable electrodes, an AD8232 signal conditioning module, an Arduino Nano 33 BLE microcontroller, and a LiPo battery in a compact enclosure. The algorithm processes ECG signals through EMD-based denoising, R-peak detection, feature extraction to compute HCM Index~1 and HCM Index~2, and dual-threshold classification. A digitization confound analysis confirmed that the classification is driven by physiological cardiac features rather than data acquisition artifacts, and a simulated device acquisition chain analysis demonstrated that the wearable hardware's signal path preserves the features required by the classification algorithm.

\subsection{Limitations}

Several limitations should be noted. First, the algorithm was validated on existing ECG databases rather than device-acquired data from HCM and LVH patients. While the simulated device acquisition chain analysis confirms that the device's signal path preserves the classification features, end-to-end validation with physical device recordings from confirmed HCM and LVH patients remains to be completed. Second, a potential confound arises from differences in data acquisition: LVH recordings were digital (10-second, 500~Hz), while HCM recordings were digitized from paper printouts (2.5 seconds per lead). The digitization confound analysis presented in Sec.~\ref{sec:confound} provides evidence that the classification is not driven by digitization artifacts; however, complete elimination of this confound requires prospective data collection with a uniform acquisition protocol. Third, while the LOOCV results demonstrate that the classification thresholds are not overfit to the evaluation data, the confidence intervals for sensitivity are wide due to the small HCM sample size ($n = 29$), and larger prospective cohorts are needed to obtain more precise estimates of diagnostic performance. Fourth, the device's 40~Hz low-pass filter bandwidth is narrower than clinical-grade ECG systems, which may limit the capture of high-frequency QRS notching that could provide additional diagnostic value; this trade-off was accepted to reduce noise in the portable form factor.

\subsection{Future Work}

Future work will focus on: (1) prospective validation using data collected directly with the wearable device from both HCM and LVH patients to eliminate potential confounds from data source heterogeneity; (2) expanding the HCM dataset through multi-center recruitment to narrow the confidence intervals on sensitivity estimates and improve the robustness of threshold optimization; (3) developing an embedded, standalone device with onboard classification that eliminates the need for a laptop, including implementation of HIPAA-compliant data encryption and secure storage; (4) investigating the addition of Lead II or other electrode configurations to capture complementary diagnostic information; and (5) multi-center clinical trials across diverse patient populations, including subgroups with athlete's heart, to validate generalizability.

%%%%%%%%%%%%%%%%%%%%%%%%%%%%%%%%%%%%%%%%%%%%%%%%%%%%%%%%%%%%%%%%%%%%%%%%%%%%%%%%%
%%  ACKNOWLEDGMENT
%%%%%%%%%%%%%%%%%%%%%%%%%%%%%%%%%%%%%%%%%%%%%%%%%%%%%%%%%%%%%%%%%%%%%%%%%%%%%%%%%
\section*{Acknowledgment}

The authors would like to acknowledge the support from Dr.\ Lin Wang (Cardiologist in Greenvale, New York) for her clinical guidance and valuable discussions on the clinical utility of the proposed screening approach.

\section*{Ethics Statement}

This study utilized de-identified ECG data from the publicly available PhysioNet database and de-identified paper ECG records provided by clinical collaborators. As no identifiable patient information was collected or used, and all data were retrospective and de-identified, this study was determined to be exempt from Institutional Review Board (IRB) review under the Common Rule (45 CFR 46.104). The simulated device acquisition chain analysis used only the same de-identified PhysioNet data and required no additional ethical approval.

%%%%%%%%%%%%%%%%%%%%%%%%%%%%%%%%%%%%%%%%%%%%%%%%%%%%%%%%%%%%%%%%%%%%%%%%%%%%%%%%%
%%  NOMENCLATURE
%%%%%%%%%%%%%%%%%%%%%%%%%%%%%%%%%%%%%%%%%%%%%%%%%%%%%%%%%%%%%%%%%%%%%%%%%%%%%%%%%

\begin{nomenclature}

\entry{ADC}{analog-to-digital converter}
\entry{BLE}{Bluetooth Low Energy}
\entry{CI}{confidence interval}
\entry{CMR}{cardiovascular magnetic resonance}
\entry{ECG}{electrocardiogram}
\entry{EMD}{Empirical Mode Decomposition}
\entry{HCM}{Hypertrophic Cardiomyopathy}
\entry{IMF}{Intrinsic Mode Function}
\entry{LA}{Left Arm electrode position}
\entry{LiPo}{lithium polymer (battery)}
\entry{LOOCV}{leave-one-out cross-validation}
\entry{LVH}{Left Ventricular Hypertrophy}
\entry{$n$}{number of heartbeats in a patient recording}
\entry{$N$}{total number of patients in the dataset}
\entry{PSD}{power spectral density}
\entry{RA}{Right Arm electrode position}
\entry{RL}{Right Leg electrode position (ground)}
\entry{SCD}{sudden cardiac death}
\entry{SF}{sampling frequency (Hz)}
\entry{SNR}{signal-to-noise ratio}

\EntryHeading{Symbols}
\entry{$I_{1,i}$}{HCM Index~1 for the $i$-th heartbeat}
\entry{$I_{2,i}$}{HCM Index~2 for the $i$-th heartbeat}
\entry{$\bar{I}_1$}{mean HCM Index~1 across all heartbeats for a patient}
\entry{$\bar{I}_2$}{mean HCM Index~2 across all heartbeats for a patient}
\entry{$\tau_1$}{classification threshold for HCM Index~1}
\entry{$\tau_2$}{classification threshold for HCM Index~2}

\end{nomenclature}

%%%%%%%%%%%%%%%%%%%%%%%%%%%%%%%%%%%%%%%%%%%%%%%%%%%%%%%%%%%%%%%%%%%%%%%%%%%%%%%%%
%%  BIBLIOGRAPHY
%%%%%%%%%%%%%%%%%%%%%%%%%%%%%%%%%%%%%%%%%%%%%%%%%%%%%%%%%%%%%%%%%%%%%%%%%%%%%%%%%

\bibliographystyle{asmejour}
\bibliography{asmejour-sample}

\end{document}